\documentstyle[12pt,epsf,a4]{article}

\newcommand{\AmS}{{\protect\the\textfont2
  A\kern-.1667em\lower.5ex\hbox{M}\kern-.125emS}}
\newcommand{\bq}{\begin{equation}}
\newcommand{\eq}{\end{equation}}
\newcommand{\beq}  {\begin{eqnarray}}
\newcommand{\eeq}  {\end{eqnarray}}
\newcommand{\rG}   {{\rm GUT}}
\newcommand{\MG}   {{\ifmmode M_\rG         \else $M_\rG$          \fi}}
\newcommand{\mb}   {{\ifmmode m_{b}         \else $m_{b}$          \fi}}
\newcommand{\mt}   {{\ifmmode m_{t}         \else $m_{t}$          \fi}}
\newcommand{\agut} {{\ifmmode \alpha_\rG    \else $\alpha_\rG$     \fi}}
\newcommand{\mgut} {{\ifmmode M_\rG         \else $M_\rG$          \fi}}
\newcommand{\mze}  {{\ifmmode m_0           \else $m_0$            \fi}}
\newcommand{\mha}  {{\ifmmode m_{1/2}       \else $m_{1/2}$        \fi}}
\newcommand{\tb}   {{\ifmmode \tan\beta     \else $\tan\beta$      \fi}}
\newcommand{\mz}   {{\ifmmode M_{Z}         \else $M_{Z}$          \fi}}
\newcommand{\ai}   {{\ifmmode \alpha_i      \else $\alpha_i$       \fi}}
\newcommand{\aii}  {{\ifmmode \alpha_i^{-1} \else $\alpha_i^{-1}$  \fi}}
\newcommand{\MSb}  {{\ifmmode \overline{\rm MS} \else
                             $\overline{\rm MS}$                   \fi}}
\newcommand{\DRb}  {{\ifmmode \overline{\rm DR} \else
      $\overline{\rm DR}$                   \fi}}
\newcommand{\DRbar}{{\ifmmode \overline{DR} \else $ \overline{DR}$ \fi}}
\newcommand{\msusy}{{\ifmmode M_{SUSY}      \else $M_{SUSY}$       \fi}}
\newcommand{\tal}  {{\ifmmode \tilde{\alpha} \else $\tilde{\alpha}$ \fi}}

\newcommand{\sws}  {{\ifmmode \;\sin^2\theta_W
                     \else    $\;\sin^{2}\theta_{W}$               \fi}}
\newcommand{\cws}  {{\ifmmode \;\cos^2\theta_W  
                     \else    $\;\cos^{2}\theta_{W}$               \fi}}
\newcommand{\sw}   {{\ifmmode\;\sin\theta_W\else $\sin\theta_{W}$  \fi}}
\newcommand{\cw}   {{\ifmmode\;\cos\theta_W\else $\;\cos\theta_{W}$\fi}}
\newcommand{\tw}   {{\ifmmode\;\tan\theta_W\else $\;\tan\theta_{W}$\fi}}
\newcommand{\bsg}  {{\ifmmode b\rightarrow s\gamma
                     \else $b\rightarrow s\gamma$ \fi}}
\newcommand{\Bbsg}  {{\ifmmode {\cal{BR}}(\b\rightarrow s\gamma)
                     \else ${\cal{BR}}(b\rightarrow s\gamma)$ \fi}}

\newcommand{\rPL}  {{\rm Planck}}
\newcommand{\mplanck} {{\ifmmode M_\rPL         \else $M_\rPL$          \fi}}
\newcommand{\rST}  {{\rm SO(10)}}
\newcommand{\msoten} {{\ifmmode M_\rST         \else $M_\rST$          \fi}}

\def\be{\begin{equation}}
\def\ee{\end{equation}}
\def\bea{\begin{eqnarray}}
\def\eea{\end{eqnarray}}
\def\rG{{\rm GUT}}

\def\MG{M_\rG}

\newcommand{\ba}   {\begin{array}}
\newcommand{\ea}   {\end{array}}

\newcommand{\lnf}  {{\ifmmode \Lambda^{(N_f)} \else $\Lambda^{(N_f)}$\fi}}
\newcommand{\ms}   {{\ifmmode \overline{MS} \else $\overline{MS}$\fi}}
\newcommand{\dr}   {{\ifmmode \overline{DR} \else $\overline{DR}$\fi}}
\newcommand{\lms}  {{\ifmmode \Lambda^{(5)}_{\overline{MS}}
                            \else $\Lambda^{(5)}_{\overline{MS}}$\fi}}
\newcommand{\lam}  {{\ifmmode \Lambda \else $\Lambda$\fi}}
\newcommand{\gev}  {{\ifmmode {\rm GeV} \else ${\rm GeV}$\fi}}
\newcommand{\gevc} {{\ifmmode {\rm GeV/c^2} \else ${\rm GeV/c^2}$\fi}}
\newcommand{\tev}  {{\ifmmode {\rm TeV} \else ${\rm TeV}$\fi}}
\newcommand{\tevc} {{\ifmmode {\rm TeV/c^2} \else ${\rm TeV/c^2}$\fi}}
\newcommand{\lp}   {{\ifmmode L^+  \else $L^+$\fi}}
\newcommand{\lm}   {{\ifmmode L^-  \else $L^-$\fi}}
\newcommand{\mlp}  {{\ifmmode M(L^-)  \else $M(L^-)$\fi}}
\newcommand{\mlz}  {{\ifmmode M(L^0)  \else $M(L^0)$\fi}}
\newcommand{\lz}   {{\ifmmode L^0     \else $L^0$\fi}}
\newcommand{\ev}   {{\ifmmode GeV/c^2       else $GeV/c^2$\fi}}
\newcommand{\tri}  {{\ifmmode \triangleup  \else $\triangleup$\fi}}
\newcommand{\unl}  {{\ifmmode U_{lL^0}  \else $U_{lL^0}$\fi}}
\newcommand{\gL}   {{\ifmmode g_L  \else $g_{L}$\fi}}
\newcommand{\gR}   {{\ifmmode g_R  \else $g_{R}$\fi}}
\newcommand{\gumu} {{\ifmmode \gamma^{\mu}  \else $\gamma^{\mu}$\fi}}
\newcommand{\gunu} {{\ifmmode \gamma^{\nu}  \else $\gamma^{\nu}$\fi}}
\newcommand{\gdmu} {{\ifmmode \gamma_{\mu}  \else $\gamma_{\mu}$\fi}}
\newcommand{\gdnu} {{\ifmmode \gamma_{\nu}  \else $\gamma_{\nu}$\fi}}
\newcommand{\stw}  {{\ifmmode\sin^2\theta_W  \else $\sin^{2}\theta_{W}$\fi}}
\newcommand{\qq}   {{\ifmmode q\overline{q} \else $q\overline{q}$\fi}}
\newcommand{\lR}   {{\ifmmode l_R  \else $l_R$\fi}}
\newcommand{\lL}   {{\ifmmode l_L  \else $l_L$\fi}}
\newcommand{\nt}   {{\ifmmode \nu_{\tau} \else $\nu_{\tau}$\fi}}
\newcommand{\nuR}  {{\ifmmode \nu_R  \else $\nu_R$\fi}}
\newcommand{\nuL}  {{\ifmmode \nu_L  \else $\nu_L$\fi}}
\newcommand{\qR}   {{\ifmmode g_R  \else $q_R$\fi}}
\newcommand{\qL}   {{\ifmmode q_L  \else $q_L$\fi}}
\newcommand{\qRp}  {{\ifmmode q_R'  \else $q_{R}$'\fi}}
\newcommand{\qLp}  {{\ifmmode q_L'  \else $q_{L}$'\fi}}
\newcommand{\est}{{\ifmmode e^{\bf \ast}  \else $e^{\bf \ast}$\fi}}
\newcommand{\lst}{{\ifmmode l^{\bf \ast}  \else $l^{\bf \ast}$\fi}}
\newcommand{\must}{{\ifmmode \mu^{\bf \ast}  \else $\mu^{\bf \ast}$\fi}}
\newcommand{\taust}{{\ifmmode \tau^{\bf \ast}  \else $\tau^{\bf \ast}$
\fi}}
\newcommand{\pperp}{{\ifmmode p_t  \else $p_t$\fi}}
\newcommand{\et}{{\ifmmode E_t  \else $E_t$\fi}}
\newcommand{\xt}{{\ifmmode x_t  \else $x_t$\fi}}
\newcommand{\smumu}{{\ifmmode \sigma_{\mu\mu}  \else $\sigma_{\mu\mu}$
\fi}}
\newcommand{\eg}{{\ifmmode e\gamma  \else $e\gamma$\fi}}
\newcommand{\epem}{{\ifmmode e^+e^-  \else $e^+e^-$\fi}}
\newcommand{\lplm}{{\ifmmode L^+L^-  \else $L^+L^-$\fi}}
\newcommand{\pp}{{\ifmmode p\overline p  \else $p\overline p$\fi}}
\newcommand{\llz}{{\ifmmode L^0\overline{L}^0 \else
$L^0\overline{L}^0$\fi}}
\newcommand{\epemt}{{\ifmmode e^+e^- \to  \else $e^+e^- \to$\fi}}
\newcommand{\eb}{{\ifmmode E_{beam}  \else $E_{beam}$\fi}}
\newcommand{\ip}{{\ifmmode pb^{-1}  \else $pb^{-1}$\fi}}
\newcommand{\upm}{{\ifmmode ^{\pm}  \else $^{\pm}$\fi}}
\newcommand{\de}{{\ifmmode ^{\circ}  \else $^{\circ}$ \fi}}
\newcommand{\appr}{{\ifmmode \sim \else $\sim$ \fi}}
\newcommand{\corresp}{{\ifmmode \stackrel{\wedge}{=}
                      \else   $\stackrel{\wedge}{=}$ \fi}}
\newcommand{\sqrts}{{\ifmmode \sqrt{s} \else $\sqrt{s}$\fi}}
\newcommand{\zz}{{\ifmmode Z^0  \else $Z^0$\fi}}
\newcommand{\mzs}{{\ifmmode M_{Z}^2  \else $M_{Z}^2$\fi}}
\newcommand{\mw}{{\ifmmode M_{W}  \else $M_{W}$\fi}}
\newcommand{\mws}{{\ifmmode M_{W}^2  \else $M_{W}^2$\fi}}
\newcommand{\mh}{{\ifmmode M_{Higgs}  \else $M_{Higgs}$\fi}}
\newcommand{\gt}{{\ifmmode \Gamma_{tot} \else $\Gamma_{tot}$\fi}}
\newcommand{\msusys}{{\ifmmode M_{SUSY}^2  \else $M_{SUSY}^2$\fi}}
\newcommand{\su}{{\ifmmode SU(3)_C\otimes\- SU(2)_L\otimes\- U(1)_Y  \else $SU(3)_C\otimes SU(2)_L\otimes U(1)_Y$\fi}}
\newcommand{\suthree}{{\ifmmode SU(3)_C  \else $SU(3)_C$\fi}}
\newcommand{\sutwo}{{\ifmmode  SU(2)_L\otimes U(1)_Y \else $SU(2)_L\otimes U(1)_Y$\fi}}
\newcommand{\taup} {{\ifmmode \tau_{proton} \else $\tau_{proton}$\fi}}
\newcommand{\mguts}{{\ifmmode M_{GUT}^2  \else $M_{GUT}^2$\fi}}
\newcommand{\mts} {{\ifmmode m_{t}^2    \else $m_{t}^2$\fi}}

\newcommand{\mtau}{{\ifmmode m_{\tau}  \else $m_{\tau}$\fi}}
\newcommand{\dpp}{{\ifmmode \delta_{pert} \else $\delta_{pert}$\fi}}
\newcommand{\dnp}{{\ifmmode\delta_{non-pert}\else$\delta_{non-pert}$\fi}}
\newcommand{\dew}{{\ifmmode \delta_{\rm EW}\else $\delta_{\rm EW}$\fi}}
\newcommand{\rt}{{\ifmmode R_{\tau}  \else
                 $R_{\tau} $\fi}}
\newcommand{\rz}{{\ifmmode R_{Z}  \else
                 $R_{Z} $\fi}}

\newcommand{\swb}{{\ifmmode \sin^2\theta_{\overline{MS}}
                     \else $\sin^2\theta_{\overline{MS}}$\fi}}
\newcommand{\cwb}{{\ifmmode \cos^2\theta_{\overline{MS}}
                     \else $\cos^2\theta_{\overline{MS}}$\fi}}
\def\ai{\alpha_i}
\def\aii{\alpha_i^{-1}}

\def\rG{{\rm GUT}}
\def\rt{{\rm threshold}}

\def\MG{M_\rG}

\def\DRbar{{\overline{DR}}}

\newcommand{\LL}{{\ifmmode {\cal L} \else ${\cal L}$\fi}}
\newcommand{\hz}{{\ifmmode {\rm Hz} \else ${\rm Hz}$\fi}}
\newcommand{\khz}{{\ifmmode {\rm kHz} \else ${\rm kHz}$\fi}}
\newcommand{\mhz}{{\ifmmode {\rm mHz} \else ${\rm mHz}$\fi}}
\newcommand{\as}{{\ifmmode \alpha_s  \else $\alpha_s$\fi}}
\newcommand{\asmz}{{\ifmmode \alpha_s(M_Z) \else $\alpha_s(M_Z)$\fi}}
\newcommand{\astau}{{\ifmmode \alpha_s(M_{\tau})
                       \else $\alpha_s(M_{\tau})$\fi}}
\newcommand{\ca}{{\ifmmode C_a  \else $C_a$\fi}}
\newcommand{\tf}{{\ifmmode T_{\mbox{\scriptsize Fermion}}
           \else $T_{\mbox{\scriptsize Fermion}}$\fi}}
\newcommand{\ts}{{\ifmmode T_{\mbox{\scriptsize Scalar}}
           \else $T_{\mbox{\scriptsize Scalar}}$ \fi}}
\newcommand{\mhiggs}{{\ifmmode M_{\mbox{\scriptsize Higgs}}
           \else $M_{\mbox{\scriptsize Higgs}}$\fi}}
\newcommand{\mthres}{{\ifmmode M_{\mbox{\scriptsize threshold}}
           \else $M_{\mbox{\scriptsize threshold}}$ \fi}}
\newcommand{\msbar}{{\ifmmode \overline{MS} \else $\overline{MS}$\fi}}
\newcommand{\drbar}{{\ifmmode \overline{DR} \else $\overline{DR}$\fi}}
\newcommand{\lamms}{{\ifmmode \Lambda_{\overline{MS}}
                       \else $\Lambda_{\overline{MS}}$\fi}}
\newcommand{\rr}{{{\ifmmode {\cal R}_2 }\else ${\cal R}_2 $\fi}}
\newcommand{\rrr}{{{\ifmmode {\cal R}_3 }\else ${\cal R}_3 $\fi}}
\newcommand{\rrrr}{{{\ifmmode {\cal R}_4 }\else ${\cal R}_4 $\fi}}
\newcommand{\jdd}{{{\ifmmode {\cal D}_2 }\else ${\cal D}_2 $\fi}}
\newcommand{\jddd}{{{\ifmmode {\cal D}_3 }\else ${\cal D}_3 $\fi}}
\newcommand{\jdddd}{{{\ifmmode {\cal D}_4 }\else ${\cal D}_4 $\fi}}
\newcommand{\rrre}{{{\ifmmode {\cal R}_3^{E0}}\else ${\cal R}_3^{E0}$\fi}}
\newcommand{\rrrp}{{{\ifmmode {\cal R}_3^P}\else ${\cal R}_3^P$\fi}}
\newcommand{\jdde}{{{\ifmmode {\cal D}_2^{E0}}\else ${\cal D}_2^{E0} $\fi}}
\newcommand{\jddp}{{{\ifmmode {\cal D}_2^P}\else ${\cal D}_2^P$\fi}}
\newcommand{\ycut}{{{\ifmmode y_{cut} }\else $y_{cut}$\fi}}
\newcommand{\ymin}{{{\ifmmode y_{min} }\else $y_{min}$\fi}}
\newcommand{\sph}{{{\ifmmode {\cal S} }\else ${\cal S} $\fi}}
\newcommand{\apl}{{{\ifmmode {\cal A} }\else ${\cal A} $\fi}}
\newcommand{\thr}{{{\ifmmode {\cal T} }\else ${\cal T} $\fi}}
\newcommand{\obl}{{{\ifmmode {\cal O} }\else ${\cal O} $\fi}}
\newcommand{\cpa}{{{\ifmmode {\cal C} }\else ${\cal C} $\fi}}
\newcommand{\eec}{{{\ifmmode {\cal E}{\cal E}{\cal C} }\else
${\cal E}{\cal E}{\cal C}$\fi}}
\newcommand{\aeec}{{{\ifmmode {\cal A}{\cal E}{\cal E}{\cal C} }
\else ${\cal A}{\cal E}{\cal E}{\cal C}$\fi}}
\newcommand{\hjm}{{\ifmmode {\bf M^2_{high}}
                   \else   ${\bf M^2_{high}}$\fi}}
\newcommand{\ljm}{{\ifmmode {\bf M^2_{low}}
                   \else   ${\bf M^2_{low}}$\fi}}
\newcommand{\djm}{{\ifmmode {\bf M^2_{diff}}
                   \else   ${\bf M^2_{diff}}$\fi}}
\newcommand{\hjmt}{{\ifmmode {\bf M({\cal T})^2_{high}}
                    \else   ${\bf M({\cal T})^2_{high}}$\fi}}
\newcommand{\ljmt}{{\ifmmode {\bf M({\cal T})^2_{low}}
                    \else   ${\bf M({\cal T})^2_{low}}$\fi}}
\newcommand{\djmt}{{\ifmmode {\bf M({\cal T})^2_{diff}}
                    \else   ${\bf M({\cal T})^2_{diff}}$\fi}}
\newcommand{\djr}{{{\ifmmode {\bf {\cal D}_2}\else ${\bf {cal D}_2}$\fi}}}
\newcommand{\ma}{{{\ifmmode {\bf {\cal M}_{Major}}
\else ${\bf {\cal M}_{Major}}$\fi}}}
\newcommand{\mi}{{{\ifmmode {\bf {\cal M}_{Minor}}
\else ${\bf {\cal M}_{Minor}}$\fi}}}

\newcommand{\ha}{{{\ifmmode {\frac{1}{2}}\else ${\frac{1}{2}}$\fi}}}

\begin{document}

\begin{titlepage}

\begin{flushright}
IEKP-KA/98-08 \\[3mm]
%{\tt hep-ph/9805378}
\end{flushright}

\vspace{2cm}

\begin{center}
  {\large\bf  Higgs Limit and $b\to s\gamma$ Constraints \\[3mm]
            in Minimal Supersymmetry} \\[1cm]

  {\bf W.~de Boer, H.-J.~Grimm,} \\[5mm]
  {\it Institut f\"ur Experimentelle Kernphysik, University of Karlsruhe \\
       Postfach 6980, D-76128 Karlsruhe, Germany} \\[1cm]

  {\bf A.V.~Gladyshev, D.I.~Kazakov} \\[5mm]

{\it Bogoliubov Laboratory of Theoretical Physics,
Joint Institute for Nuclear Research, \\
141 980 Dubna, Moscow Region, Russian Federation}

\end{center}

\vspace{2cm}

\abstract{
New limits on the Higgs mass from LEP and new calculations on the radiative 
(penguin) decay of the $b$-quark into $s\gamma$ 
restrict the parameter 
space of the Constrained Minimal Supersymmetric Standard Model (CMSSM).

We find that for the low $\tb$ scenario only one sign of the Higgs mixing 
parameter is allowed, while  the high $\tb$ scenario is practically
excluded, if one requires all sparticles to be below 1 TeV and imposes
radiative electroweak symmetry breaking as well as gauge and Yukawa coupling
unification. For squarks between 1 and 2 TeV  high $\tb$ scenarios are allowed.
We consider especially a new high $\tb=64$ scenario with triple unification
of all Yukawa couplings of the third generation, which shows
an infrared fixed point behaviour.

The upper limit on the  mass of the lightest Higgs in the low (high) $\tb$ 
scenarios is $97\pm6~(120\pm2)$ GeV,
where the largest error originates predominantly from the 
uncertainty in the top mass.

}

\end{titlepage}

%
% -----------------------------------------------------------------
%
\section{Introduction}

Recently several new results have  appeared,
which have a considerable impact on the allowed parameter space of the Constrained
Minimal Supersymmetric Model (CMSSM). As constraints we consider gauge unification and
radiative electroweak symmetry breaking  assuming the supergravity inspired scenario
of the soft breaking of the SUSY masses with common scalar and gaugino masses
at the GUT scale\cite{rev}.
The  additional reduction of the parameter space by 
relic density, $b-\tau$ Yukawa coupling unification and the new Higgs limits
from LEP as well as the present $\bsg$ rate is calculated.

The most significant new results are:
\begin{itemize}
\item
the SM Higgs mass limit is now 89.3 GeV at 95\% C.L. 
for the SM Higgs~\cite{moriondh}, which  corresponds to the CMSSM case too, 
since  all the other Higgses are heavy and decouple in the CMSSM limit.
\item
Higher order   corrections to the $\bsg$ calculation became available \cite{misiak,marciano,neubert}.
As experimental value we use the combined value 
from the CLEO Coll.\cite{cleo}  and  ALEPH Coll.\cite{aleph}
\Bbsg=$(2.54\pm0.56)\cdot10^{-4}$, which is 
consistent with the SM expectation: 
\Bbsg=$(3.6\pm0.3)\cdot10^{-4}.$ This SM  value was calculated for 
$\as=0.122$, which is the best value from the
$Z^0$ data of the four LEP experiments\cite{jer}.
Since the calculations are usually quoted as the ratio of the $\bsg$ 
rate and the
semileptonic decay rate  of the b-quark, 
one needs for absolute predictions  
${\cal{BR}}(b\rightarrow X_c ~e\overline{\nu})$,
 for which we use  $10.5\pm0.4$\%\cite{blnvalue}.

In a recent paper the theoretical uncertainties on the photon spectrum
are studied  in next-to-leading order as well\cite{neubert}.
They could increase the CLEO value by as much as 14\%. 
Such a correction improves the
agreement between data and the SM prediction. Given the present 
 uncertainty of 22\%
in the published data, this correction has not  been 
taken into account,
so we stick to the published values and added in the analysis
the theoretical   uncertainty  from the SM prediction, the semileptonic
branching ratio uncertainty and the experimental errors in quadrature.

\end{itemize}

Our statistical analysis, in which the probability for each point of
parameter space is determined by a $\chi^2$ fit, has been described previously~\cite{wezp}.
The masses of the third generation and the couplings constants are taken 
from LEP results presented at the Jerusalem Conf. \cite{jer}, except
for the top mass from the combined data from D0 and CDF, which was updated: 
$m_t=173.9\pm5.2$ GeV at Moriond\cite{moriond}.

Since the common GUT masses $\mze$ and $\mha$ for the spin 0 and spin 1/2
particles are 
strongly correlated, we perform the fit for all combinations of  $m_0$ and 
$m_{1/2}$ between 100 GeV and 1 TeV in steps of 100 GeV.

%
% ------------------------------------------------------------------
%
\section{Results}

\subsection{Constraints from the top and bottom mass}
\label{sec:su5}

The requirement of bottom-tau Yukawa coupling unification strongly
restricts the possible solutions in the $\mt$ versus $\tb$ plane, as
discussed in our previous paper\cite{wezp}.

\begin{figure}[ht]
\vspace{-1cm}
    \begin{center}
    \leavevmode
    \epsfxsize=10cm
    \epsffile{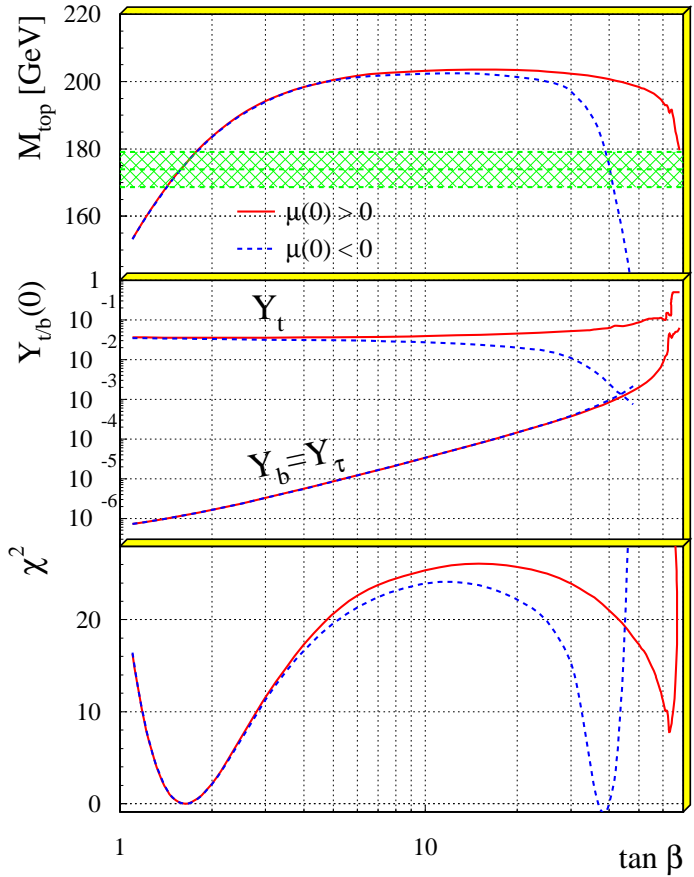}
\end{center}
\vspace{-0.5cm}
  \caption[]{\label{f2}The top quark mass as function of $\tb$ (top) 
      for values of $\mze,\mha~\approx 1 $ TeV.  The curve is hardly 
      changed for lower SUSY masses.
      The middle part shows the corresponding values of the Yukawa
      couplings at the GUT scale and the lower part the
      $\chi^2$ values.
      If the top constraint ($\mt=174\pm5$, horizontal band) 
      is not applied, all values of $\tb$ between 1.2 and 
      70 are allowed,
       but if the top mass 
      is constrained to the experimental value, only the regions
      $\tb=1.65\pm0.3$, $ \tb \sim 35$, and $ \tb \sim 64$  are 
      allowed.
      }
 \end{figure}

Only three values of $\tb$ give an
acceptable $\chi^2$ fit for $m_t=174$ GeV, as shown in the bottom part of
fig.~\ref{f2}. Clearly, the value $\tb=64$ gives a worse $\chi^2$
than the values at 35 and 1.65, which is mainly due to
the fact that this solution cannot fit simultaneously the
top and bottom mass with a single $\tb$ if $b-\tau$ unification is required
( either $\mt$ or  $m_b$ too large by about 3$\sigma$).

Therefore it was not considered in our previous paper\cite{wezp}. 
We consider it here,  since the solution with $\tb\approx 35$ does
not give a better $\chi^2$  anymore in the combined fit of all constraints
because of  the smaller errors from the gauge couplings, 
$m_t$ and $\bsg$.

\begin{figure}[tb]
    \begin{center}
\vspace*{-3cm}
    \leavevmode
    \epsfxsize=9cm
  \epsffile{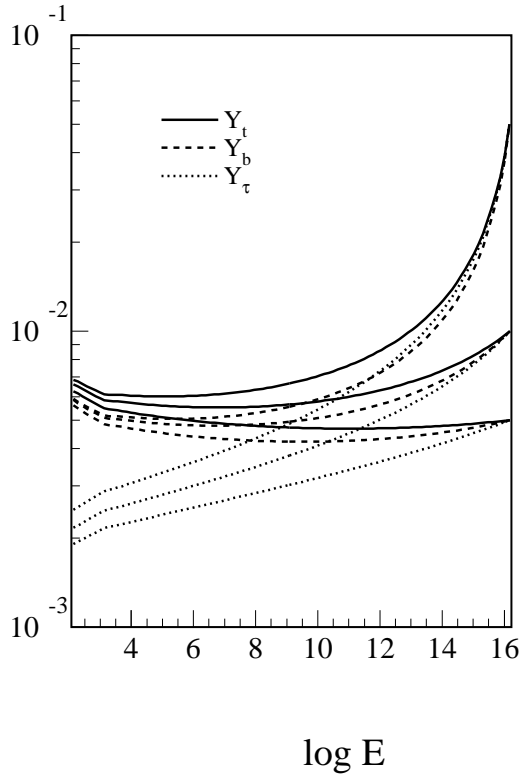}
\vspace*{-1.5cm}
\end{center}
  \caption[]{\label{f3} The running of the Yukawa couplings in case 
$Y_t=Y_b=Y_\tau$ at the GUT scale ($SO(10)$ type solution). One can clearly 
 see the approach to the three different fixed points, i.e. the low energy
value is largely independent of the GUT scale value. Consequently the
GUT scale values can be choosen to be equal (triple unification).
The fixed point values at low energy  yield
 correct masses for   bottom and tau for $\tb\approx 64$; the fixed
point of the top mass yields 
$m_t^2=(4\pi)^2 Y_t v^2 \sin^2\beta$=184 GeV, 
which is about $2\sigma$ above
the experimental value.  
 The values of the Yukawa couplings at 
the GUT scale are shown in the maximum allowed range: higher values
of $Y_t$ tend to produce negative Higgs masses, while lower 
values yield a too low top mass .
 }
 \end{figure}

Furthermore, this solution has the unique
feature of a possible triple Yukawa unification:  
all 3 Yukawa couplings are 
driven to an approximate fixed point, which yield approximately the 
correct masses of the leptons and quarks of the third generation. 
Since the mass deviation are at the $3\sigma$ level, we consider it
here, especially since the solution with $\tb\approx 35$ does
not give a better $\chi^2$  with the smaller errors from the gauge couplings, 
$m_t$ and $\bsg$.
 The fixed point is reached for a large range of initial values of the
Yukawa couplings at the GUT scale, as shown in fig. \ref{f3}.

The difference between the two solutions at high $\tb$ 
corresponding to  opposite signs of $\mu$ stems from finite loop
corrections to the bottom quark mass  involving squark-gluino and stop-chargino
loops.  These corrections are small for low \tb
solutions, but can become as large as 10-20\% 
for the high \tb values\cite{wezp},
since the dominant  corrections are proportional to $\mu\tan\beta$.
Consequently they change sign for different signs of $\mu$.
As shown in the middle part of fig. \ref{f2} the Yukawa couplings become of the same
order of magnitude at large $\tb$. For $\tb=64$ $Y_b$ varies so rapidly that it is easy to find a solution where all Yukawa 
couplings are equal, but large at the GUT scale.
The low energy value is then approximately given by a 
fixed point solution for each of 
the 3 Yukawa couplings, i.e. the solutions of the RGE are 
insensitive to the
starting values at the GUT scale.

Radiative electroweak symmetry breaking is only possible in that case, if one assumes
non-universal values for the mass parameters in the Higgs potential. 

As will be discussed below, the \bsg rate is also sensitive to the
sign of $\mu$ and it  turns out to be difficult to have solutions,
where both $b-\tau$ unification and the $\bsg$ rate are correctly described.

%
%----------------------fixed points- htb -----------------cd
%
%-----------------bsgamma
%
%
\begin{figure}[htb]
%  \vspace*{-1.5cm}
  \begin{center}
    \leavevmode 
%\rotate[r]{
    \epsfxsize=13.0cm
    \epsffile{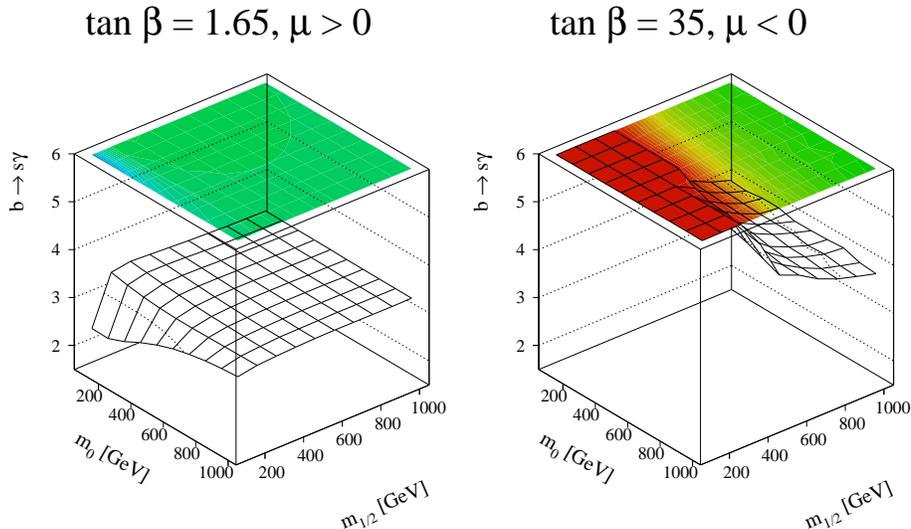}
%}
  \end{center}
  \vspace*{-0.8cm}
  \caption[]{The predicted $b\to s\gamma$ branching ratio 
            for low and high $\tb$ in units of 10$^4$ as function of $\mze$ and $\mha$. 
            At low $\tb$ the prediction is close to the SM value (3.5) in the whole plane,
            while at high $\tb$ it can be enhanced by a large factor for light 
            sparticles (mainly charginos).\label{f5} 
}
\end{figure}

% 
%--------------- chi^2
%
  \begin{figure}[htb]
%    \vspace{-0.51cm}
\begin{center}
    \leavevmode
    \epsfxsize=10.0cm
    \epsffile{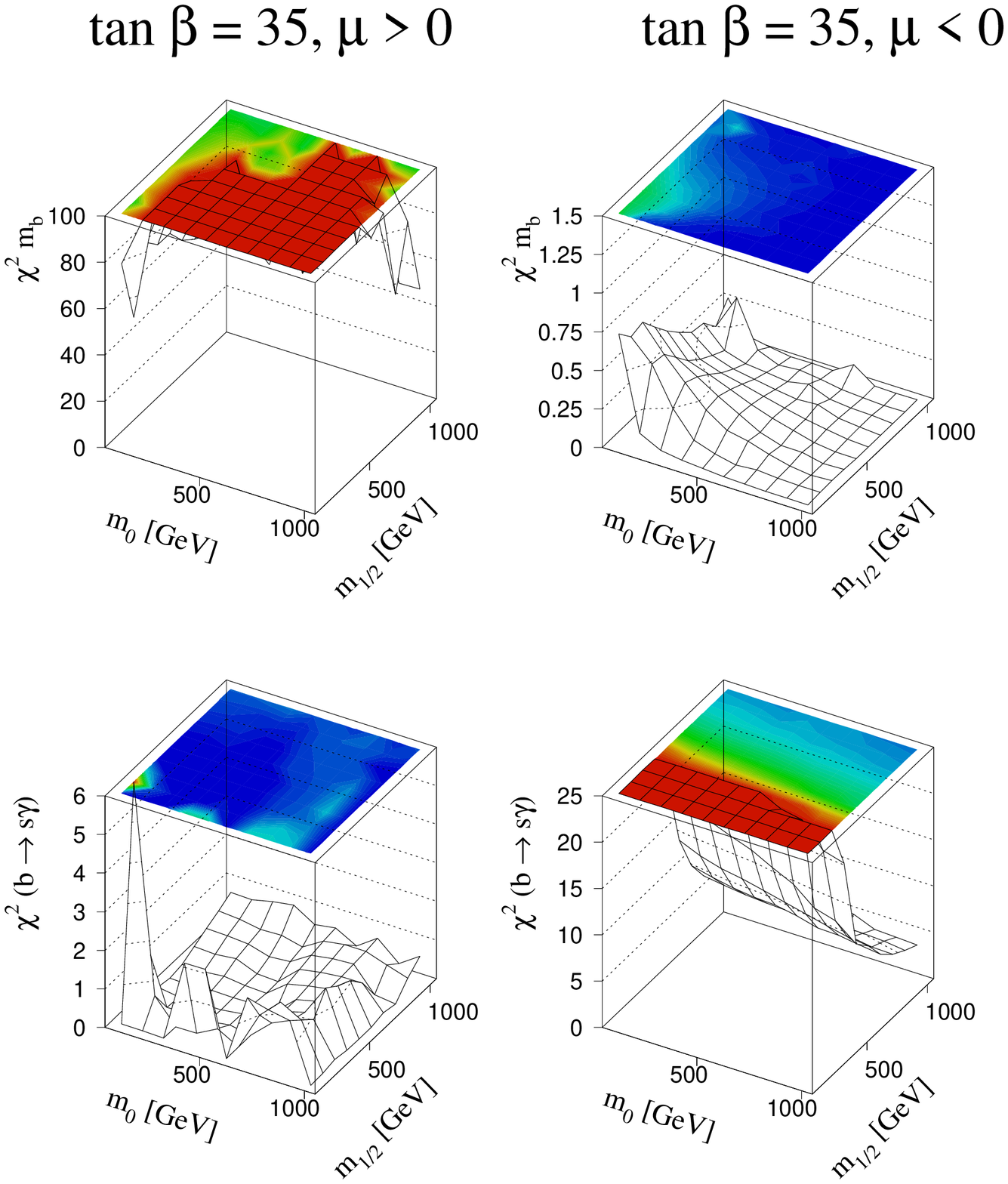}
\end{center}
%\vspace{-0.8cm}
\caption[]{The  $\chi^2$ contributions from the bottom mass  and $\bsg$  
for different signs of $\mu$ in the high $\tb$ solution.
Note  that $\bsg$ is perfect for $\mu<0$, while the
bottom mass is perfect  for $\mu>0$. Requiring both yields a low $\chi^2$
only for heavy sparticles ($\mha>650$ GeV). 
}\label{f9}
\end{figure}

% ---------------------------------------------------------------------
%
\subsection{\bsg decay rate}

Recently the Next-to-Leading Order (NLO) QCD contributions to the
$\bsg$ decay rate 
have been calculated in the SM\cite{misiak}.
They increase the SM value by about 10\%. 
The NLO electroweak (EW) contributions
decrease the result by a few \%\cite{neubert,marciano}.
In NLO the large theoretical uncertainty from the scale
dependence is now reduced to the 10\% level 
(see e.g. ref. \cite{neubert} for a recent summary on 
the discussion of 
the uncertainties).
The NLO corrections  can be simulated 
in the lowest order calculation by choosing a renormalization scale 
$\mu\approx0.6m_b$.
Since the NLO have not yet been calculated for the MSSM,
we will stick to the first order calculations, but choose
$\mu=0.6m_b$. Since the 
NLO corrections are small compared with
 the experimental uncertainty of 22\%, we 
expect that with the present large experimental uncertainties
 the NLO corrections in the MSSM will not
change the conclusions significantly.

%The complete lowest order calculations were published by Bertolini 
%et al~\cite{bsg4}. We do not use the approximate codes
%from Barbieri and Giudice\cite{bsg3} or Oshimo\cite{Oshimo},
%since the first completely neglects the intergenerational mixing, while
%the last one overestimates it.
The \bsg branching ratio is very sensitive to the presence of SUSY
particles. Their effect has been calculated by Bertolini
et al~\cite{bsg4} where a complete but cumbersome matrix notation for the amplitudes
was used. We use the  more explicit
 expressions by Barbieri and Giudice~\cite{bsg3}. However, in the latter paper
 the intergenerational mixing in squark mass matrices is omitted, which is
not  justified in all regions of parameter space:
 we find it  can change the charged current amplitude with charginos 
in the loop by $\approx 10\%$ or more. So the usual assumption
that intergenerational mixing predominantly introduces only the usually negligible
flavour changing neutral current with gluinos and down-type quarks in the loop,
is not valid.
Therefore we extended the explicit formulae of Barbieri and Giudice
with the intergenerational mixing, thus obtaining an efficient code, which
is important for numerical $\chi^2$ analysis.

%Therefore we took the dominant intergenerational mixing between
%the second and third generation into account assuming that  at the GUT scale
% the Yukawa matrices are diagonal and the mixing is generated 
%by the CKM matrix multiplied by radiative corrections for the 
% flavour changing charged currents, while  the intergenerational mixing for the FCNC
%is generated through radiative corrections alone. 
%The complete formula will be published separetely\cite{wdbzp}.

At low \tb the \bsg rate is close to its SM value for most of the plane 
(see fig. \ref{f5}).  The charginos and/or the  charged Higgses are only light 
enough at small values of $m_0$ and $m_{1/2}$ to contribute significantly. 
%The trilinear couplings were found to play a negligible role for low \tb. 

 However, for large $\tb$   the \bsg rate strongly depends on the
particle spectrum and can be larger than the SM value by factors 3 or more,
as shown on the right hand side of fig. \ref{f5}.
%The individual contributions 
%are shown in fig.~\ref{f6} as function of the sparticle masses, i.e.
% as function of $\mze$ and $\mha$.
%The MSSM contributions are 
%of the same order of magnitude as the SM contributions, 
%which are -0.00636 and -0.01225 for the W-t loop and operator mixing term, 
%respectively.
%%------------------------------------------------------------------------
The chargino contribution, which is proportional to $\mu\ \tb$,
changes sign for different signs of $\mu$. Through interference with
the large operator mixing term it can lower or increase the total \bsg rate.
The gluino contribution is noticeable only if $m_{1/2}$ is small. However, in this
region the charginos are light too, so the chargino contribution  is always larger 
by an order of magnitude. 

For negative $\mu$ values the \bsg rate is usually too large, except if all sparticles
are very heavy and don't contribute (see fig. \ref{f5}, right hand side).
For positive $\mu$ one can reach the values of \bsg rate close to the 
experimental value in most of the parameter region, as is apparent from
the $\chi^2$ distribution in the left hand bottom side of fig. \ref{f9}.
However it is difficult
to fit simultaneously  $m_b$, if the Yukawa coupling $Y_b$ is determined  
from $m_\tau$ via $b-\tau$ unification (see left hand top part of fig. \ref{f9}).
Note that for this sign of $\mu$ 
the $m_b$ mass is not just slightly wrong, for which threshold corrections
could be blamed, but for our case off by more than $7 \sigma$, 
where we used the conservative Particle Data Book
errors for the running mass of $m_b=4.2\pm0.15$ GeV. Recent determinations
from Ypsilon mass spectra quote a 50\% lower error\cite{melnikov}.

On the right hand side of fig. \ref{f9} the sign of $\mu$ is reversed.
In this case $m_b$ can be fitted very well, but $\bsg$ cannot be fitted
 for $\mha<600$ GeV.
The reason is that in this case the negative chargino loop contribution 
$\propto \mu \tb$ yields a strong positive contribution through the
interference with the negative operator mixing term, so \Bbsg comes out too large,
if the charginos are light. On the contrary,  
the large $m_b$ contributions from $\tilde{g}-\tilde{q}$ and 
$\tilde{\chi}^\pm - \tilde{t}$ loops proportional to $\mu\tb$
now lower $m_b$ sufficiently.
% They  become 
%of the order of 10-20\%~\cite{wezp}.  In order to obtain $m_b(M_Z)$ as low as 
%2.84 GeV,  these corrections have to be negative, thus requiring $\mu$ to be 
%negative, but then $\bsg$ is too large in a large region, as shown in the
%right hand bottom side of fig. \ref{f9}; 
% $m_b$ is fitted very well in that case (right hand top side of \ref{f9}.

Our conclusions may seem to contradict  those of refs.\cite{raby-new} and
\cite{baer}. In ref. \cite{raby-new} the discrepancy between the left bottom
and left top of fig. \ref{f9} is solved by minimizing the corrections to $m_b$
proportional to $\as\mu\tb$, which can be done by a small value of $\mu$ and
a small value of $\as$. Giving up exact gauge unification by  using 
larger errors on the gauge couplings and allowing for 
an  offset of a few \%
 for the strong coupling $\as$ at the GUT scale, 
they find acceptable fits.
%The reason is that in ref.\cite{raby-new} the authors allow
%an arbitrary offset of a few \% for
%the strong coupling $\as$ at the GUT scale. Such non-unification
%may originate from threshold corrections  at the GUT 
%scale~\cite{lukas}.
%Assuming in addition orders of magnitude larger errors in the electroweak
%coupling constants than given by LEP and the Particle Data Book, which results in 
%a larger uncertainty in $\alpha_{GUT}$,
%They can fit  both $m_b$ and $\bsg$ simultaneously for $\mu>0$ 
% at the price of non-unification for the gauge couplings and $\as(M_Z)\approx 0.114$ 
%for heavy squarks.
%This value is well below the LEP value of $0.122\pm0.003$.

Our $\bsg$ calculations  are  in reasonable agreement with the results from ref. \cite{baer}
for both signs of $\mu$. However, since they do not impose $b-\tau $ unification, their
preferred solution is $\mu>0$, which corresponds to the left bottom corner of fig. \ref{f9}
and ignoring the left top corner.
% 
%--------------- chi^2
%
  \begin{figure}[htb]
%    \vspace{-1.51cm}
\begin{center}
    \leavevmode
    \epsfxsize=14.0cm
    \epsffile{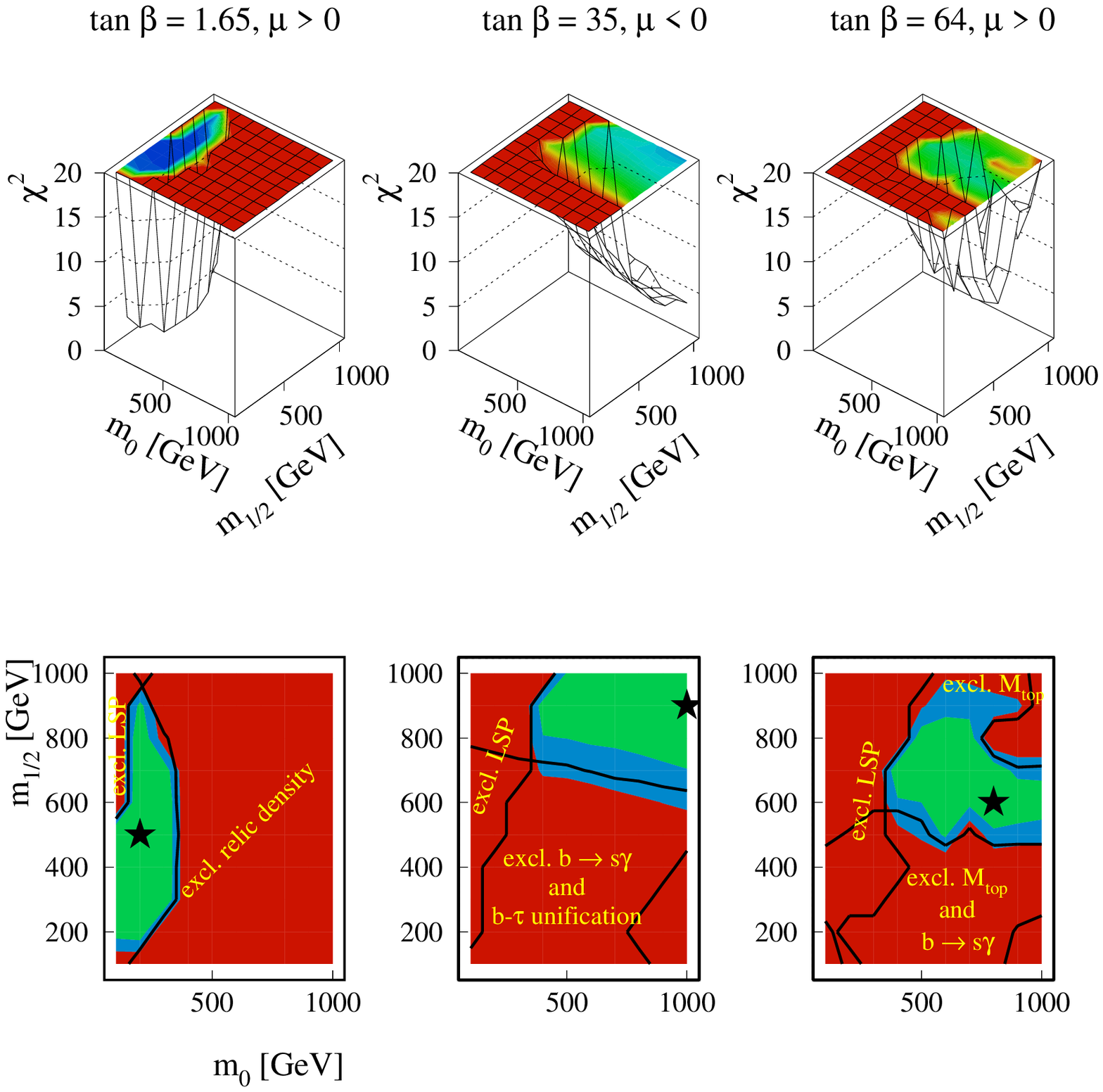}
\end{center}
\vspace{-0.8cm}
\caption[]{\label{f8}The  $\chi^2$-distribution for 
  the low and high $\tb$ solutions.  The different shades
   in the projections  indicate
  steps of $\Delta\chi^2 = 4$, so basically only the light shaded
  region is allowed.
      The stars indicate the optimum solution. 
      Contours enclose domains excluded by the particular
      constraints used in the analysis.
}
\end{figure}
%  
%--------------- chi^2
%
  \begin{figure}[htb]
   \vspace*{-.8cm}
\begin{center}
    \leavevmode
    \epsfxsize=10.0cm
    \epsffile{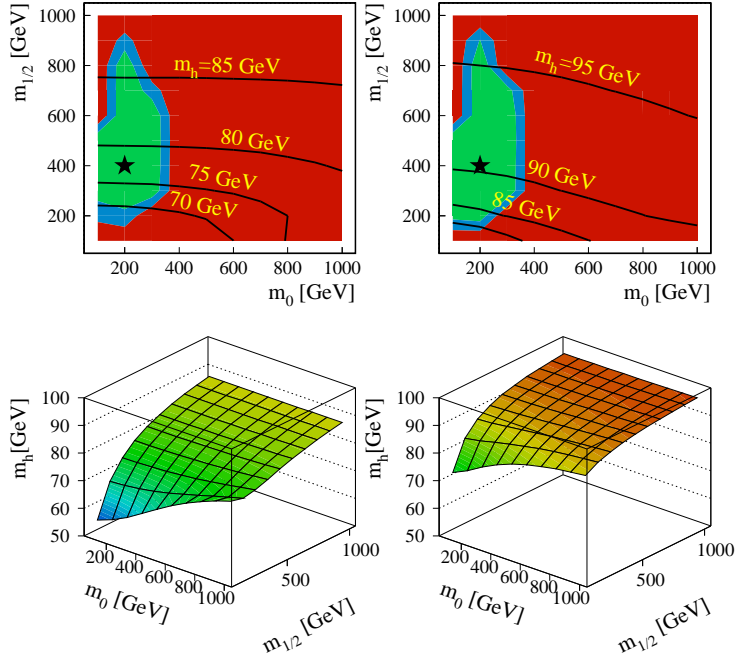}
\end{center}
\vspace{-1.0cm}
\caption[]{\label{f10}Contours of the  Higgs mass (solid lines) 
  in the $m_0,m_{1/2}$ plane (above)  and the Higgs masses (below) for both 
    signs  of $\mu$  for the low $\tb$ solution $\tb=1.65$ for $M_t=175$ GeV.
}
\end{figure}
%
%----------------- mh vs. mtop
%
%
\begin{figure}[htb]
  %\vspace*{-1.05cm}
  \begin{center}
    \leavevmode
    \epsfxsize=10.0cm
    \epsffile{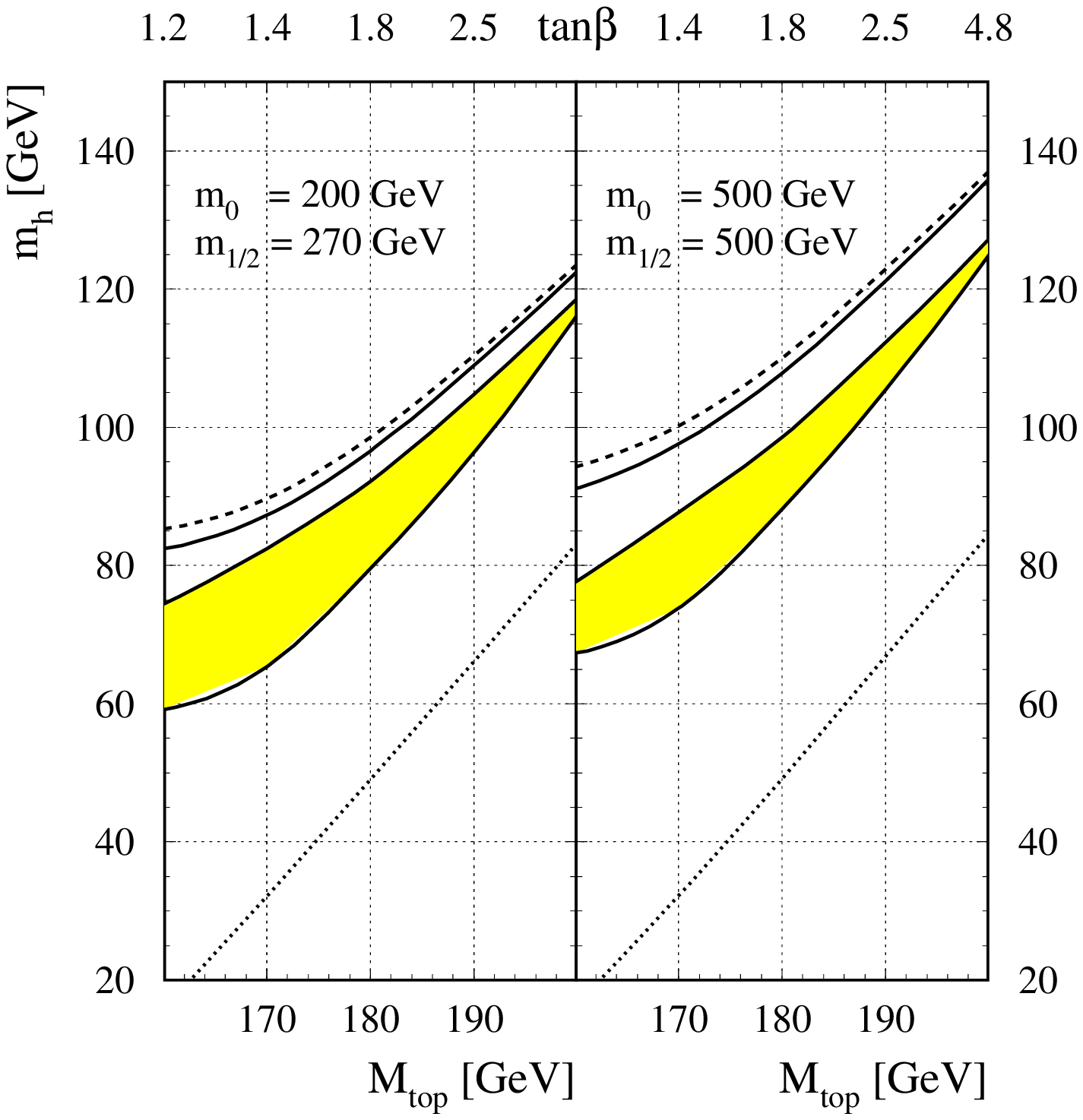}
  \end{center}
  \vspace*{-0.3cm}
  \caption[]{\label{f12}The mass of the lightest CP-even Higgs as 
    function of the top mass at Born level (dotted lines), 
    including complete one-loop contributions of all particles
    (dashed lines). Two-loop contributions\cite{ll}
 reduce the one-loop
    corrections\cite{bekhiggs} significantly as shown by the shaded area
    (the upper boundary corresponds to $\mu>0$, the lower one
    to $\mu<0$). The solid line just
    below the dashed line is the one-loop prediction 
    from the third generation only,
    which apparently gives the main contribution. The upper scale
    indicates the value of $\tb$, as calculated from the top mass
    by the requirement of $b-\tau$-unification.
    }
\end{figure}
\subsection{Excluded parameter space and the mass spectrum}

In fig.~\ref{f8} the total $\chi^2$ distribution is shown as a
function of $\mze$ and $\mha$ for the three values of $\tb$ determined
above from $b-\tau$ unification. 
The areas at low $m0$ and high $\mha$ are excluded by the LSP constraint,
since in that case the lightest $\tilde{\tau}$ can become the LSP. If R-parity
is conserved, a charged LSP is not allowed, since the vacuum would be filled
with charged relics from the Big Bang.

The relic density constraint excludes $\mze>350 $ GeV in case of small $\tb$,
as discussed previously\cite{wezp}. For large
$\tb$ the Higgsino mixture of the LSP allows a fast enough decay via s-channel
$Z^0$ exchange, so in that case the requirement  $\Omega h^2 \le 1$ is easily fulfilled.

As discussed in the previous section the combined requirements
of  the correct \bsg rate and $b-\tau$ unification 
exclude quite a region of parameter space for large $\tb$, as shown by the contours
in the lower part.

One observes  $\chi^2$ minima at $\mze,\mha$ around (200,500),
(1000,900), and (500,500) for the different $\tb$ values,
respectively, as indicated by the stars.

%The mass spectra corresponding to the minima are given in  
%table~\ref{t2} together with the values of partial amplitudes 
%contributing to the \bsg decay rate. 

Note that the squarks and gluinos are typically above
1 TeV for the high \tb solutions. Furthermore, the minimal $\chi^2$
values are not excellent for high \tb: for 
$\tb=64$ $\chi^2_{min}$=6.1 from the fitted top mass
 ($\mt=189$ GeV), while for $\tb=35$
  $\chi^2_{min}$=4.3 from \bsg. All other $\chi^2$ contributions are 
negligible.
For   $\tb=1.65$ $\chi^2_{min}=1.7$, basically
from the \Bbsg constraint alone. 

Apart from the heavy spectra for large \tb, 
 one has the problem that
the Born term Higgs masses are strongly negative, as can be anticipated
from the fast running of the soft mass terms of the two Higgs doublets, 
$m_1^2$ and $m_2^2$, which receive negative radiative corrections proportional to 
the Yukawa couplings. For large $\tb$ both $Y_t$ and $Y_b$ are large (see fig. \ref{f2}).
Then the pseudoscalar Higgs mass, 
 ($m_1^2+m_2^2$ at the Born level), and the lightest Higgs get negative masses at the
Born level, which can only become positive by large  radiative corrections.
In this case it is extremely important to include the full 1-loop radiative
corrections, i.e. one has to include both  the sparticles and particles in the loops.
For the full 1-loop corrections we used the formulae from ref. \cite{bekhiggs}.
For the lightest Higgs the dominant second order corrections from ref. \cite{ll}
have been included as well.

For low \tb the present Higgs limits severely constrains 
the parameter space, as can be seen from fig. \ref{f10}, which
shows the  excluded regions in the ($m_0,m_{1/2})$ plane for  
different signs of $\mu$.
As mentioned in the introduction
the SM Higgs limit of 89.3 GeV is valid for the low $\tb$ 
scenario ($\tb<4$)
of the MSSM too. As is apparent from fig. \ref{f10} 
this limit clearly rules out
the $\mu<0$ solution,  in agreement with other studies\cite{abel}. 
However, this figure assumes $\mt=175$ GeV.
 The top dependence on the Higgs
mass is slightly steeper than  linear in this range (see 
fig. \ref{f12}). Adding about one $\sigma$ to the top mass, 
i.e. $\mt=180$ GeV,
implies that for the contours in fig. \ref{f10} one should add 6 GeV to the numbers shown.
Even in this case the $\mu<0$ solution is excluded for a large 
region of parameter
space. Only the small allowed  region with
$\mha>700$ GeV is not yet excluded for $\mt=180$ GeV. Note that in this region
the squarks are well above  1 TeV, so in this case the cancellation of the 
quadratic divergencies in the Higgs masses, which is only perfect if sparticles and
particles have the same masses,  starts to become worrying again.

%For squarks masses around 1 TeV ($\mze=500,\mha=500$ GeV) the Higgs masses
%are $92\pm6$ ($81\pm6$) GeV for  $\mu>0~ (\mu<0)$.
%As shown on the left side of fig. \ref{f8}, such large $\mze$ values are excluded
%by the relic density constraint. 
%For typical squark masses in the
%preferred region of fig. \ref{f10} ($\mze=200,\mha=500$) the Higgs masses are
%$92\pm6$ ($81\pm6$) GeV for $\mu>0 ~(\mu<0)$, repectively, as can be read from fig. \ref{f10}.
In summary, if all squarks are required to be below 1 TeV the $\mu<0$ solutions
are excluded for the low $\tb$ scenario, even if the uncertainty from the top mass
is taken into account.
For $\mu>0$ practically the whole plane  is allowed, except for the left bottom corner
shown on the top right hand side of fig. \ref{f10}.

The  upper limit for the mass of the lightest Higgs is reached for heavy squarks, but it saturates
quickly, as is apparent from the bottom row in fig. \ref{f10}.
For $\mze=1000,\mha=1000$, which corresponds to  squarks masses of about 2 TeV\footnote{Explicit
analytical expressions for the sparticle masses as function of the SUSY parameters
can be found in ref. \cite{wekzp}.}, one finds
for the upper limit on the Higgs mass in the low $\tb$ scenario:
$$m_h^{max}=97\pm6~{\rm GeV},$$
 where the error is dominated by the uncertainty from the top mass.
If one requires the squarks to be below 1 TeV, these upper limits are reduced by 4 GeV. 

 It  agrees well with the value from Casas et al.: $m_h=97\pm2$ GeV\cite{Haber}.
In both analysis  the Renormalization Group Equations are used to determine the trilinear coupling
at low energies and $\mu$ from EWSB, so the mixing in the stop sector is fixed, once
the sign of $\mu$ is choosen. Furthermore in both papers 
solutions close to the infrared fixed point are considered, which are required
in our case from the EWSB. 
Our  error on the upper limit is larger than the one from Casas et al., since
they did not consider the error on the top mass.

For high $\tb$ the upper limit on the Higgs mass in the CMSSM is:$$m_h^{max}=120\pm2~{\rm GeV}.$$
The error from the top mass is small, since the high $\tb$ fits prefer anyway
top masses around 190 GeV.

%
% --------------------------------------------------------------------- 
%
\section{Summary}

It is shown that the CMSSM can fit simultaneously the constraints from
\begin{itemize}
\item gauge couplings unification;
\item $b-\tau$ Yukawa couplings unification;
\item radiative EWSB;
\item life time of the Universe.
\end{itemize}

The present Higgs limit of 89.3 GeV  largely excludes the 
$\mu<0$ solution for the low $\tb$  scenario.  The $\mu>0$ solution is just within 
reach of the coming LEP II runs above 189 GeV, since the upper limit
on the Higgs mass is $97\pm6$ GeV. Here the central value of 97 GeV was calculated
for $\mt=175$ GeV and $m_h=103$ GeV is reached for $\mt=180$ GeV.
If one requires the squarks to be below 1 TeV, these upper limits are reduced by 4 GeV. 

The maximum value of the lightest  Higgs mass in the high $\tb$ scenarios is $120\pm2$ GeV, 
which is beyond the reach of LEP II. However, the high $\tb$ solutions 
have serious 
finetuning problems, because of the fact that all Yukawa couplings are large, 
which cause the Higgs masses at tree level to be  strongly negative (typically -1 TeV),
so the radiative corrections have to be positive and very large to offset this large
negative "starting" value. 
 Furthermore, the solutions with minimal $\chi^2$ require  
squarks above 1 TeV,  which have an additional finetuning problem
because of the non-cancellation of the quadratic divergencies to the Higgs masses.
These quadratic corrections only cancel perfectly, if sparticles and particles have
the same mass.

\section*{Note added in proof.}
After completing this work, first results on the NLO corrections
to the $\bsg$ rate in the MSSM became available\cite{ciuchini}.
They change the amplitudes by a similar amount as the SM amplitudes,
but due to the large cancellations in particlular regions
of parameter space, the effect can be large sometimes.
This will be studied in a future paper.
We thank G.F. Giudice for a discussion on this point.

\end{document}